\documentclass[aps,showpacs,prl,superscriptaddress]{revtex4} 

\usepackage{sidecap}
\usepackage{ulem}
\usepackage{epsfig}
\usepackage{amsmath,amssymb,amsthm}
\usepackage{graphicx}
\usepackage{bm}
\usepackage{color}

\DeclareMathAlphabet{\mathpzc}{OT1}{pzc}{m}{it}

\begin{document}

\renewcommand{\textfraction}{0.00}

% Useful macros:

\newcommand{\vAi}{{\cal A}_{i_1\cdots i_n}} 
\newcommand{\vAim}{{\cal A}_{i_1\cdots i_{n-1}}} 
\newcommand{\vAbi}{\bar{\cal A}^{i_1\cdots i_n}}
\newcommand{\vAbim}{\bar{\cal A}^{i_1\cdots i_{n-1}}}
\newcommand{\htS}{\hat{S}} 
\newcommand{\htR}{\hat{R}}
\newcommand{\htB}{\hat{B}} 
\newcommand{\htD}{\hat{D}}
\newcommand{\htV}{\hat{V}} 
\newcommand{\cT}{{\cal T}} 
\newcommand{\cM}{{\cal M}} 
\newcommand{\cMs}{{\cal M}^*}
\newcommand{\vk}{\vec{\mathbf{k}}}
\newcommand{\bk}{\bm{k}}
\newcommand{\kt}{\bm{k}_\perp}
\newcommand{\kp}{k_\perp}
\newcommand{\km}{k_\mathrm{max}}
\newcommand{\vl}{\vec{\mathbf{l}}}
\newcommand{\bl}{\bm{l}}
\newcommand{\bK}{\bm{K}} 
\newcommand{\bb}{\bm{b}} 
\newcommand{\qm}{q_\mathrm{max}}
\newcommand{\vp}{\vec{\mathbf{p}}}
\newcommand{\bp}{\bm{p}} 
\newcommand{\vq}{\vec{\mathbf{q}}}
\newcommand{\bq}{\bm{q}} 
\newcommand{\qt}{\bm{q}_\perp}
\newcommand{\qp}{q_\perp}
\newcommand{\bQ}{\bm{Q}}
\newcommand{\vx}{\vec{\mathbf{x}}}
\newcommand{\bx}{\bm{x}}
\newcommand{\tr}{{{\rm Tr\,}}} 
\newcommand{\bc}{\textcolor{blue}}

\newcommand{\beq}{\begin{equation}}
\newcommand{\eeq}[1]{\label{#1} \end{equation}} 
\newcommand{\ee}{\end{equation}}
\newcommand{\bea}{\begin{eqnarray}} 
\newcommand{\eea}{\end{eqnarray}}
\newcommand{\beqar}{\begin{eqnarray}} 
\newcommand{\eeqar}[1]{\label{#1}\end{eqnarray}}
 
\newcommand{\half}{{\textstyle\frac{1}{2}}} 
\newcommand{\ben}{\begin{enumerate}} 
\newcommand{\een}{\end{enumerate}}
\newcommand{\bit}{\begin{itemize}} 
\newcommand{\eit}{\end{itemize}}
\newcommand{\ec}{\end{center}}
\newcommand{\bra}[1]{\langle {#1}|}
\newcommand{\ket}[1]{|{#1}\rangle}
\newcommand{\norm}[2]{\langle{#1}|{#2}\rangle}
\newcommand{\brac}[3]{\langle{#1}|{#2}|{#3}\rangle} 
\newcommand{\hilb}{{\cal H}} 
\newcommand{\pleft}{\stackrel{\leftarrow}{\partial}}
\newcommand{\pright}{\stackrel{\rightarrow}{\partial}}
%%%%%%%%%%%%%%%%%%%%%%%%%%%%%%%%%%%%%%%%%%%%%%%%%%%%%%%%%%%%%%%%%%%%%%%%%%%%%%

\title{RHIC and LHC jet suppression in non-central collisions}

\date{\today}
 
\author{Magdalena Djordjevic}
\affiliation{Institute of Physics Belgrade, University of Belgrade, Serbia}
\author{Marko Djordjevic}
\affiliation{Faculty of Biology, University of Belgrade, Serbia}
\author{Bojana Blagojevic}
\affiliation{Institute of Physics Belgrade, University of Belgrade, Serbia}

\begin{abstract} 
Understanding properties of QCD matter created in ultra-relativistic heavy-ion collisions is a major goal of RHIC and LHC experiments. An excellent tool to study these properties is jet suppression of light and heavy flavor observables. Utilizing this tool requires accurate suppression predictions for different experiments, probes and experimental conditions, and their unbiased comparison with experimental data. With this goal, we here extend our dynamical energy loss formalism towards generating predictions for non-central collisions; the formalism takes into account both radiative and collisional energy loss, dynamical (as opposed to static) scattering centers, finite magnetic mass, running coupling and uses no free parameters in comparison with experimental data. Specifically, we here generate predictions for all available centrality ranges, for both LHC and RHIC experiments, and for four different probes (charged hadrons, neutral pions, D mesons and non-prompt $J/\psi$). We obtain a very good agreement with all available non-central data, and also generate predictions for suppression measurements that will soon become available. Finally, we discuss implications of the obtained good agreement with experimental data with different medium models that are currently considered. 
\end{abstract}


\maketitle 

\section{Introduction} 
Jet suppression~\cite{Bjorken} of light and heavy observables provides an excellent tool~\cite{Brambilla,Gyulassy,DBLecture} for studying properties of QCD matter created in ultra-relativistic heavy ion collisions. Mapping these properties is also a major goal of RHIC and LHC experiments, which requires comparison of jet suppression measurements with corresponding theoretical predictions. To ensure the unbiased comparison with experimental data, it is necessary to generate predictions for different experiments, experimental probes and experimental conditions, within the same theoretical model. With a major goal of generating these predictions, we developed dynamical energy loss formalism, that {\it i)} allows treating, at the same time, both light and heavy partons, {\it ii)} is computed in dynamical QCD medium (i.e. takes into account recoil of the medium constituents), {\it iii)} includes both collisional~\cite{MD_Coll} and radiative~\cite{MD_PRC,DH_PRL} energy losses, computed within the {\it same} theoretical framework, {\it iv)} includes realistic finite size effects, i.e. the fact that experimentally created QCD medium has finite size, and that the jets are produced inside the medium, {\it v)} includes finite magnetic mass effects~\cite{MD_MagnMass}  and running coupling~\cite{MD_PLB2014}. We further integrated this formalism into numerical procedure which also includes multi-gluon  fluctuations~\cite{GLV_suppress}, path length fluctuations~\cite{WHDG}  and most up-to-date jet production~\cite{Cacciari:2012,Vitev0912} and fragmentation functions~\cite{DSS}; the procedure allows generating predictions with no free parameters used in comparison with experimental data.

We previously applied the computational procedure outlined above for generating predictions in {\it most central} collisions for a number of different probes at LHC~\cite{MD_PLB2014}. These predictions showed a very good agreement with experimental data; however, a comprehensive comparison also requires generating predictions for non-central collisions at RHIC and LHC. With this goal, we here extend the formalism towards generating predictions for different centrality ranges. We consequently generate the suppression predictions for all available centrality ranges, for both RHIC and LHC experiments and for four different probes - specifically for charged hadrons,  D mesons and non-prompt $J/\psi$ at LHC and neutral pions at RHIC. Such comprehensive comparison allows testing some of important assumptions behind our current understanding of the created QCD matter, such as ranges of validity for different medium models.

\section{Theoretical framework}
The numerical procedure for calculating jet suppression for central collisions is outlined in detail in~\cite{MD_PLB2014}. We below first briefly list the main steps in this procedure and then describe the extension of the procedure that is necessary for generating the predictions for non-central collisions:

{\it i)} Energy loss calculations: Our model takes into account both radiative and collisional contributions to jet energy loss. Specifically, the radiative energy loss calculations present a state-of-the-art extension of a well-known DGLV model~\cite{GLV,DG} towards a finite size dynamical medium~\cite{MD_PRC,DH_PRL}, finite magnetic mass~\cite{MD_MagnMass} and running coupling~\cite{MD_PLB2014}. These extensions are further discussed below: 

{\it ii)} Dynamical scattering centers: To calculate the radiative energy loss, we use finite size dynamical energy loss formalism. This formalism removes an ubiquitous assumption of static scattering centers~\cite{Gyulassy_Wang} and takes into account that the medium constituents are in reality dynamical, i.e. moving particles; similarly, the unrealistic assumption of infinite medium is also removed. Calculations of the jet energy loss in dynamical medium are done by using two-hard-thermal-loop approach. In contrast to the static energy loss, where only the electric contribution appears in the final result, both electric and magnetic contributions appear in the dynamical case. This then directly leads to the question of finite magnetic mass, which we further discuss below.

{\it iii)} Magnetic mass: In pQCD energy loss calculations - including our (initial) dynamical energy loss formalism~\cite{MD_PRC,DH_PRL} - magnetic mass is taken to be zero. However, different non-perturbative approaches suggest a non-zero magnetic mass at RHIC and LHC (see e.g.~\cite{Nakamura,Maezawa,Hart,Bak}). To address this issue, we generalized the dynamical energy loss calculations to the case of finite magnetic mass. Introducing the finite magnetic mass is described in detail in~\cite{MD_MagnMass}, where the finite magnetic mass is introduced through generalized sum-rules. 

{\it iv)} Running coupling: Introducing the running coupling is described in detail in~\cite{MD_PLB2014}. One should note that the obtained result is infrared safe and moreover of a moderate value. There is consequently no need to introduce an artificial cutoff as is commonly done elsewhere with the running coupling. 

{\it v)} Jet suppression procedure: We further integrated the energy loss formalism outlined above into a numerical procedure that includes: light and heavy flavor production~\cite{Cacciari:2012,Vitev0912}, path-length~\cite{WHDG} and multigluon~\cite{GLV_suppress} fluctuations, up-to-date fragmentation functions~\cite{DSS} for light and heavy flavor and the decay of heavy mesons to single electrons and J/$\psi$. In the calculations, as a start point we use an effective temperature of 304 MeV for 0-40$\%$ centrality Pb+Pb collisions at LHC (as extracted by ALICE~\cite{ALICE_T}) and effective  temperature of 221 MeV for 0-20$\%$ centrality Au+Au collisions at RHIC (as extracted by PHENIX~\cite{RHIC_T}). The other parameter values are specified in the next section, while the details of the procedure are provided in~\cite{MD_PLB2014}. Note that we use no free parameters in comparison with the data, i.e. all the parameters that we use correspond to standard literature values.

To extend the computational procedure outlined above to non-central collisions, we start by obtaining the path-length distributions for different centrality ranges from~\cite{Dainese}. Furthermore, we determine the temperature for each centrality region according to~\cite{GLV} $T^3 \sim \frac{\frac{dN_g}{dy}}{V} \rightarrow  T= c \left(\frac{\frac{dN_g}{dy}}{N_{part}} \right)^{1/3}$ (for more details see~\cite{MD_Temp}), 
where $\frac{dN_g}{dy}$ is gluon rapidity density, $V$ is the volume of created medium, and we take that $V \sim N_{part}$ (number of participants for a given collision). Furthermore, c is a constant for a specific system/collider energy, and $\frac{\frac{dN_g}{dy}}{N_{part}}$ is directly proportional to experimentally measured charged particle multiplicity per participant pair ($\frac{\frac{dN_{ch}}{dy}}{N_{part}/2}$), which is measured for  both RHIC~\cite{RHIC_mult} and LHC~\cite{LHC_mult} and across different centralities. The constants c can be fixed through ALICE measurement of effective temperature for $0-40\%$ centrality at 2.76 TeV Pb+Pb collisions LHC, and through PHENIX measurement of effective temperature for $0-20\%$ centrality at 200 GeV Au+Au collisions at RHIC (see above). 

\section{Numerical results} 

In this section, we concentrate on 200 GeV Au+Au collisions at RHIC and 2.76 
TeV Pb+Pb collisions at LHC, and present our suppression predictions for light 
and heavy flavor observables. We proceed by considering a QGP  with 
$n_f{\,=\,}2.5$ effective light quark flavors for RHIC and $n_f{\,=\,}3$ for 
LHC. Perturbative QCD scale is taken to be $\Lambda_{QCD}=0.2$~GeV. For the 
light quarks we assume that their mass is dominated by the thermal mass 
$M{\,=\,}\mu_E/\sqrt{6}$, where the temperature dependent Debye mass $\mu_E (T)$  is obtained from~\cite{Peshier}.  Magnetic mass $\mu_M$ is taken as $0.4 \, \mu_E < \mu_M < 0.6 \, \mu_E$~\cite{Maezawa,Nakamura,Hart,Bak}, and the gluon mass is  $m_g=\mu_E/\sqrt{2}$~\cite{DG_TM}. For the 
charm (bottom) mass we use $M{\,=\,}1.2$\,GeV ($M{\,=\,}4.75$\,GeV). Path-length distribution and temperatures for different centralities are computed according to the procedure outlined in the previous section. Parton production, fragmentation functions and decays, which are used in the numerical calculations, are specified in~\cite{MD_PLB2014}. Note that, on each 
panel of every figure, the gray region corresponds to the range of $0.4 < \mu_M/\mu_E < 0.6$, where the upper (lower) 
boundary of each band corresponds to $\mu_M/\mu_E =0.6$ ($\mu_M/\mu_E =0.4$).
%
%%%%%%%%%%%%%%%%%%%%%%%% Fig. 1 %%%%%%%%%%%%%%%%%%%%%%%%%%%%%%%%%%%%%%%%%%%%
\begin{figure*}
\epsfig{file=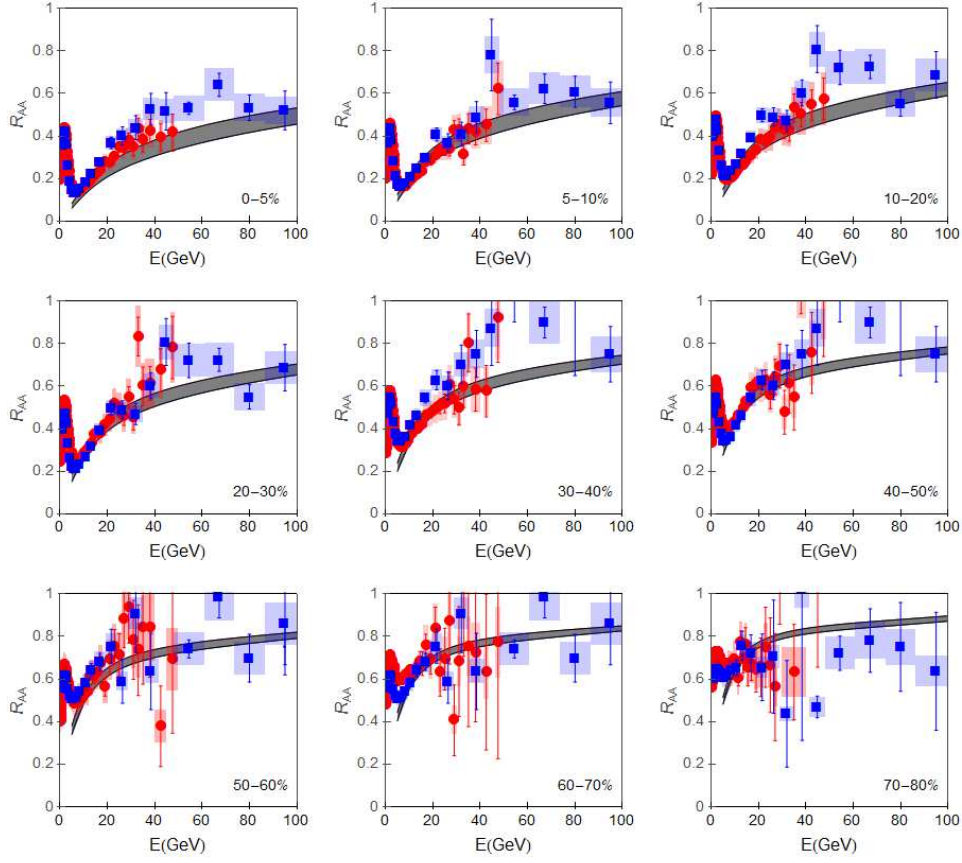,width=5in,height=4.5in,clip=5,angle=0}
%\vspace*{-0.7cm}
\caption{{\bf Theory {\it vs.} experimental data for momentum dependence of charged hadron $R_{AA}$ for different centrality bins at LHC.} The panels show the comparison of charged hadron suppression 
predictions with experimentally measured $R_{AA}$ for charged particles at 2.76 Pb+Pb collisions at LHC, for different (fixed) centrality ranges. Red circles and blue squares correspond to ALICE~\cite{ALICE_h} and CMS~\cite{CMS_h} experimental data, respectively. In the lower right corner of each panel we denote the centrality for which the data and the predictions are presented. Note that, on the third and the fourth panel, CMS data for centrality bin 10-30\% are shown. Similarly, on the fifth and the sixth panel, CMS data for centrality bin 30-50\% are shown, on the seventh and the eight panel, CMS data for centrality bin 50-70\% are shown, and on the ninth panel CMS data for 70-90\% in centrality are shown.}
\label{LightHadronRaaLHC}
%\end{minipage}
\end{figure*}
%%%%%%%%%%%%%%%%%%%%%%%%%%%%%%%%%%%%%%%%%%%%%%%%%%%%%%%%%%%%%%%%%%%%%%%%%%%%

%%%%%%%%%%%%%%%%%%%%%%%% Fig. 1 %%%%%%%%%%%%%%%%%%%%%%%%%%%%%%%%%%%%%%%%%%%%
\begin{figure*}
\epsfig{file=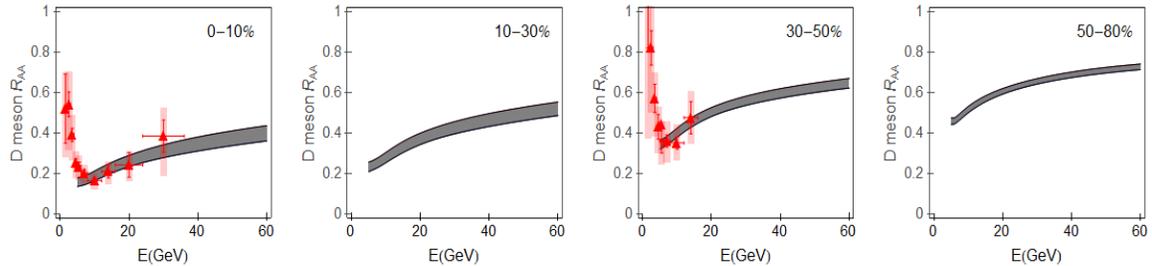,width=6in,height=1.5in,clip=5,angle=0}
\vspace*{-0.4cm}
\caption{{\bf Theory {\it vs.} experimental data for momentum dependence of D meson $R_{AA}$ for different centrality bins at LHC.} The left panel shows the comparison of D meson suppression 
predictions with D meson $R_{AA}$ at $0-7.5\%$ central 2.76 Pb+Pb collisions at LHC~\cite{ALICE_D} (the red triangles). The other three panels show the theoretical predictions for D meson $R_{AA}$ for centrality bins $10-30 \%$, $30-50 \%$ and $50-80 \%$, respectively. In the third panel ($30-50\%$ centrality region), the predictions are compared with ALICE preliminary data~\cite{ALICE_QM} that recently became available.}
\label{DSuppRaaLHC}
%\end{minipage}
%\end{minipage}
\end{figure*}
%%%%%%%%%%%%%%%%%%%%%%%%%%%%%%%%%%%%%%%%%%%%%%%%%%%%%%%%%%%%%%%%%%%%%%%%%%%%

%%%%%%%%%%%%%%%%%%%%%%%% Fig. 1 %%%%%%%%%%%%%%%%%%%%%%%%%%%%%%%%%%%%%%%%%%%%
\begin{figure*}
\epsfig{file=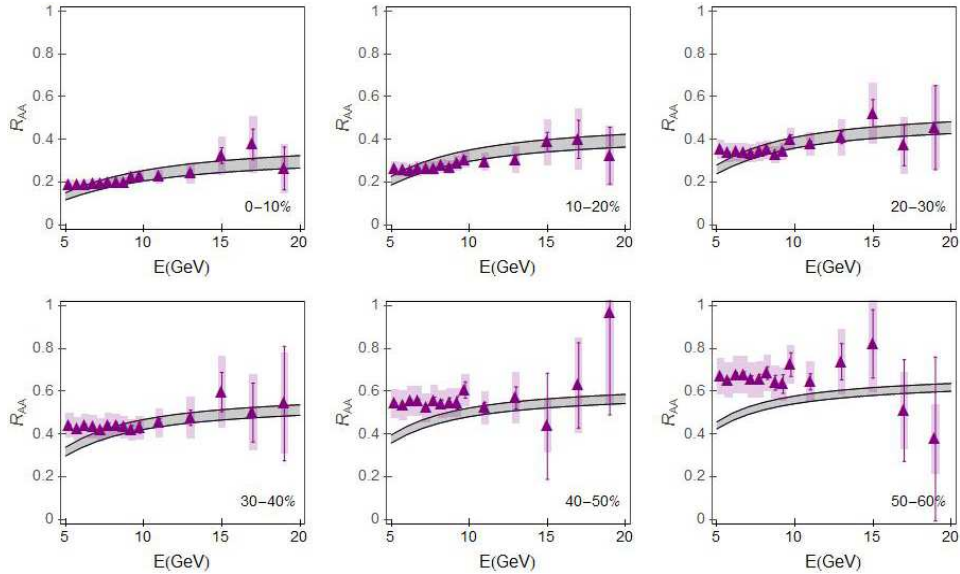,width=5in,height=3.in,clip=5,angle=0}
%\vspace*{-0.7cm}
\caption{{\bf Theory {\it vs.} experimental data for momentum dependence of neutral pion $R_{AA}$ for different centrality bins at RHIC.} The panels show the comparison of neutral pion suppression 
predictions with $R_{AA}$ for neutral pions at 200 GeV Au+Au collisions at RHIC, for different (fixed) centrality regions. Purple triangles correspond to PHENIX~\cite{PHENIX} data. In the lower right corner of each panel we denote the centrality for which the data and the predictions are presented. } 
\label{LightHadronRaaRHIC}
%\end{minipage}
%\end{minipage}
\end{figure*}
%%%%%%%%%%%%%%%%%%%%%%%%%%%%%%%%%%%%%%%%%%%%%%%%%%%%%%%%%%%%%%%%%%%%%%%%%%%%

%%%%%%%%%%%%%%%%%%%%%%%% Fig. 1 %%%%%%%%%%%%%%%%%%%%%%%%%%%%%%%%%%%%%%%%%%%%
\begin{figure*}
\epsfig{file=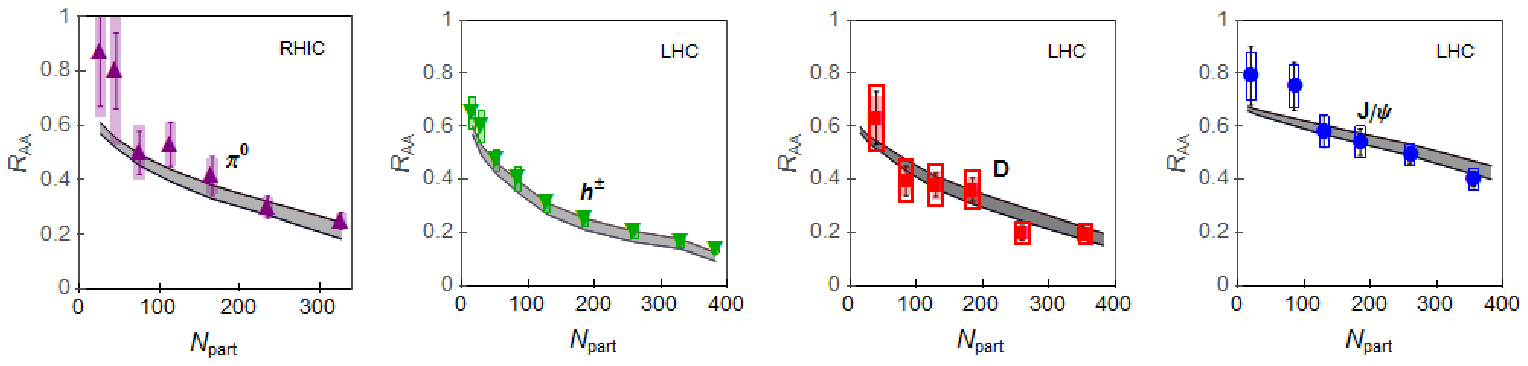,width=6in,height=1.5in,clip=5,angle=0}
\vspace*{-0.4cm}
\caption{{\bf Theory {\it vs.} experimental data for participant dependence of light and heavy flavor $R_{AA}$  at RHIC and LHC.} The first panel compares theoretical predictions with experimental data for participant dependence of $\pi^0$ $R_{AA}$~\cite{PHENIX} at 200 GeV Au+Au collisions at RHIC, where $\pi^0$ momentum is larger than 7 GeV. The second, third and fourth panel compares theoretical predictions with experimental data for participant dependence of, respectively, $h^\pm$~\cite{CH_CentDep}, D meson~\cite{D_CentDep} and non-prompt $J/\psi$~\cite{JPsi_CentDep} $R_{AA}$ at 2.76 TeV Pb+Pb collisions at LHC.  The jet momentum ranges for the second, the third and the forth panel are, respectively, 6-12GeV, 8-16GeV and 6.5-30GeV.}
\label{RaaVsCent}
%\end{minipage}
%\end{minipage}
\end{figure*}
%%%%%%%%%%%%%%%%%%%%%%%%%%%%%%%%%%%%%%%%%%%%%%%%%%%%%%%%%%%%%%%%%%%%%%%%%%%%

We start by generating predictions for momentum dependence of jet suppression at  LHC experiments, for different centrality regions, which are shown in Figs. 1 and 2. Each panel in these figures shows a fixed centrality region ($0-5\%$, $5-10\%$, $10-20\%$, etc.) and for each of these centrality regions, momentum dependence of $R_{AA}$ is shown. Figure 1 shows predictions for charged hadron $R_{AA}$ and their comparison with relevant ALICE and CMS experimental data at 2.76 TeV Pb+Pb collisions at LHC. In Fig. 2 predictions for D meson $R_{AA}$ are shown; predictions for $0-10\%$ and $30-50\%$ are compared with the available ALICE data, where a very good agreement can also be seen. Note that predictions for $30-50\%$ region were generated {\it before} the experimental data – that are now shown in the figure – became available~\cite{MD_QM}. The experimental data for the rest of the predictions (the other two panels in Fig. 2) are expected to become available soon.

In Figure 3, we show equivalent predictions as in Figs. 1 and 2, but for RHIC measurements of neutral pions at 200 GeV Au+Au collisions. Each panel shows predictions for different (fixed) centrality bin, which are compared with experimental data. Similarly as for LHC measurements, we see a very good agreement between the theoretical predictions and RHIC data. 

In Figure 4, instead of fixing the centrality ranges (as in Figs. 1-3), we fix the momentum regions and explore how $R_{AA}$ changes for different centrality values (i.e. number of participants). The predictions are generated for both RHIC and LHC experiments, and for various probes. Specifically, we compare our predictions with experimental data for neutral pions at RHIC and charged hadrons, D mesons and non-prompt $J/\psi$ at LHC. One can see that we here also obtain a robust agreement with the experimental data.

\section{Conclusions} 

We here generated suppression predictions for all available centrality ranges, for both RHIC and LHC, and for diverse experimental probes. These predictions were generated by the same theoretical formalism and within the same numerical procedure. Furthermore, all the predictions within the same experiment (i.e. within RHIC and within LHC) were generated with the same parameter set, which corresponds to standard literature values, and with no free parameters used in comparison with experimental data. We obtained an excellent agreement of the theoretical predictions with the diverse experimental measurements, for all momentum ranges larger than 10 GeV. 

The robust agreement discussed above has interesting implications for ranges of validity of different medium models, which are incorporated in different approaches to jet suppression predictions. As discussed in the Introduction, our calculations employ state-of-the-art method for energy loss calculations and numerical procedure for suppression calculations, but do not explicitly take into account the medium evolution (i.e. the evolution is taken into account through effective/average medium parameters). This is in contrast to a number of other approaches (see e.g.~\cite{BGMNQRR,BetzGyulassy,BAMPS,GNBGA}), which simplify the energy loss to a various degree, in order to more explicitly incorporate the evolving medium.  Consequently, the obtained robust agreement with the experimental data above 10 GeV strongly suggests that, for hard probes, expansion of the medium does not play a major role in explaining the jet propagation in ultra-relativistic heavy-ion collisions. We hypothesize that the reason behind this result is that hard probes have such a large amount of energy and the created medium is so short, that the hard probe propagation is only sensitive to the average properties of the created medium. This hypothesis remains to be tested in the future, and ranges of validity for different medium models have to be mapped. Such mapping can both considerably simplify theoretical predictions, and even more importantly, facilitate intuitive understanding of complex experimental data.

{\em Acknowledgments:} 
This work is supported by Marie Curie International Reintegration Grant 
within the $7^{th}$ European Community Framework Programme 
(PIRG08-GA-2010-276913) and by the Ministry of Science and Technological 
Development of the Republic of Serbia, under projects No. ON171004 and 
ON173052. We thank A. Dainese for providing part-length distributions for different centrality ranges.

\end{document}